\documentclass[journal]{IEEEtran}
\IEEEoverridecommandlockouts
\usepackage{cite}
\usepackage[colorlinks=true,linkcolor=black,anchorcolor=black,citecolor=black,filecolor=black,menucolor=black,runcolor=black,urlcolor=magenta]{hyperref}

\usepackage{amsmath,amssymb,amsfonts,mathrsfs}
\usepackage{algorithmic}
\usepackage{graphicx}
\usepackage{textcomp}
\usepackage{xcolor}
\usepackage{booktabs}
\usepackage{subeqnarray}
\usepackage{multicol}
\def\BibTeX{{\rm B\kern-.05em{\sc i\kern-.025em b}\kern-.08em
    T\kern-.1667em\lower.7ex\hbox{E}\kern-.125emX}}
\usepackage[defaultcolor=red, final]{changes}

\ifCLASSOPTIONcompsoc
\usepackage[caption=false,font=normalsize,labelfont=sf,textfont=sf]{subfig}
\else
\usepackage[caption=false,font=footnotesize]{subfig}
\fi

\newcommand{\bs}[1]{\boldsymbol{#1}}

\newcommand{\overbar}[1]{\mkern 1.5mu\overline{\mkern-1.5mu#1\mkern-1.5mu}\mkern 1.5mu}

\begin{document}
\bstctlcite{Settings}
\title{Indirect Measurement of Switch Terms of a Vector Network Analyzer with Reciprocal Devices}

\author{%
	\IEEEauthorblockN{%
		Ziad~Hatab, Michael~Ernst~Gadringer, and~Wolfgang~Bösch
	}%
	\thanks{This work was supported by the Christian Doppler Research Association and the Austrian Federal Ministry for Digital and Economic Affairs and the National Foundation for Research, Technology, and Development. The authors are with the Institute of Microwave and Photonic Engineering, and the Christian Doppler Laboratory for Technology-Guided Electronic Component Design and Characterization (TONI), Graz University of Technology, 8010 Graz, Austria (e-mail: z.hatab@tugraz.at; michael.gadringer@tugraz.at; wbosch@tugraz.at).}
\thanks{Software code and measurements are available online:\\ \url{https://github.com/ZiadHatab/vna-switch-terms}}
}%
\markboth{This work has been accepted for publication in the IEEE Microwave and Wireless Technology Letters}{}
\maketitle

\begin{abstract}
This paper presents an indirect method for measuring the switch terms of a vector network analyzer (VNA) using at least three reciprocal devices, which do not need to be characterized beforehand. This method is particularly suitable for VNAs that use a three-sampler architecture, which allows for applying first-tier calibration methods based on the error box model. The proposed method was experimentally verified by comparing directly and indirectly measured switch terms and performing a multiline thru-reflect-line (TRL) calibration.
\end{abstract}


\section{Introduction}
\label{sec:1}
\IEEEPARstart{C}{alibration} of a vector network analyzer (VNA) is crucial to remove systematic errors between the device under test (DUT) and the actual receivers of the VNA. The most common calibration method is the short-open-load-thru (SOLT) method, which is based on the 12-term error model of a two-port VNA. However, this method requires fully characterized standards. In \cite{Ferrero1992}, a modification to the SOLT method was introduced, where the thru standard is replaced with any transmissive reciprocal device, called the SOLR method. This method is based on the error box model of a two-port VNA. Other advanced self-calibration methods, including thru-reflect-line (TRL), multiline TRL, line-reflect-match (LRM), and line-reflect-reflect-match (LRRM) \cite{Engen1979, Marks1991, Eul1988, Hayden2006}, also rely on the error box model of a two-port VNA.

\begin{figure}[th!]
	\centering
	\subfloat[]{\includegraphics[width=.49\linewidth]{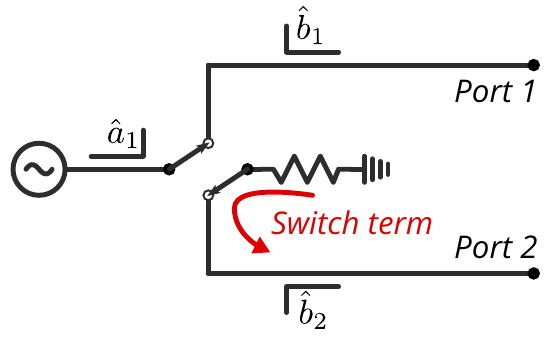}
		\label{fig:1.1a}}
	\subfloat[]{\includegraphics[width=.49\linewidth]{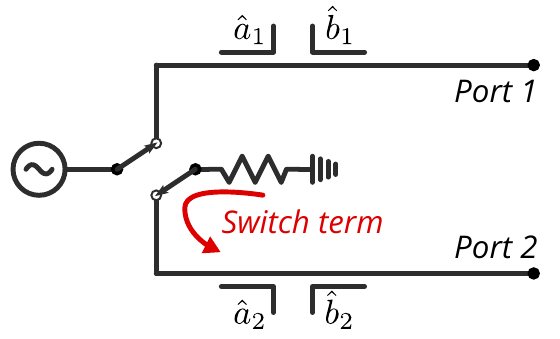}
		\label{fig:1.1b}}
	\caption{Illustration of three- (a) and four-sampler (b) architectures of a VNA. Both diagrams depict source driving in the forward direction.}
	\label{fig:1.1}
\end{figure}

A limitation of calibration methods based on the error box model is that they require a four-sampler (dual reflectometer) VNA to sample all waves, whereas the 12-term error model can still be used in the three-sampler (single reflectometer) VNA. Fig.~\ref{fig:1.1} illustrates the two architectures for a two-port VNA. The difference between the two architectures is that in the three-sampler VNA we do not sample the reflected wave of the termination load of the non-driving port. Although the termination load is generally designed to be matched, in reality, there is always some reflection that needs to be accounted for. This reflection is called the switch term. Since the ports are driven in both forward and reverse directions, there are two switch terms for the two-port configuration. For the general multiport configuration, there are $N$ switch terms, where $N$ is the number of ports. In the dual reflectometer architecture, $2N$ samplers are used, where each port has two samplers, whereas for the single reflectometer architecture, $N+1$ samplers are used, where each port has one sampler and the source has its own sampler.

Given that the termination of the non-driving ports remain constant during measurement, the switch terms introduce systematic error and only need to be measured once. These terms can be considered as part of the calibration coefficients through the conversion relationships between the 12-term and error box models \cite{Marks1997,Vandenberghe2002,Hayden2007,Jargon2018}.

For three-sampler VNAs, self-calibration based on the error box model cannot be used, as switch terms cannot be directly measured. Instead, an SOLT calibration or equivalent can be performed with known standards using the 12-term or 10-term model (ignoring crosstalk) as generalized in \cite{Ferrero2008} for multiport VNAs. Additionally, error box calibration can be performed as a second-tier after SOLT calibration \cite{Jargon1995}. However, SOLT calibration requires pre-characterized calibration standards, which contradicts the purpose of self-calibration using partially defined standards.

This paper aims to introduce a new method to indirectly measure the switch terms using at least three transmissive reciprocal devices, which do not need to be characterized beforehand. The proposed method enables the usage of error box calibration methods in three-sampler VNAs without requiring any prior first-tier calibration.

\section{Mathematical Formulation}
\label{sec:2}

\subsection{Problem statement}

In a two-port VNA, when all four waves are sampled in both driving directions, the measured S-parameters are described using the following notation \cite{Rumiantsev2008}:
\begin{equation}
	\begin{bmatrix}
		\hat{b}_{11} & \hat{b}_{12}\\
		\hat{b}_{21} & \hat{b}_{22}
	\end{bmatrix} = \bs{S}\begin{bmatrix}
	\hat{a}_{11} & \hat{a}_{12}\\
	\hat{a}_{21} & \hat{a}_{22}
	\end{bmatrix}
	\label{eq:2.1}
\end{equation}
where $\hat{a}_{ij}$ and $\hat{b}_{ij}$ represent the sampled incident and reflected waves, respectively, at port-$i$ when driven by port-$j$. 

In a three-sampler VNA, the waves $\hat{a}_{12}$ and $\hat{a}_{21}$ are not measured due to a lack of dedicated receivers. To address this, the measured incident waves in \eqref{eq:2.1} can be split into two matrices as follows:
\begin{equation}
	\begin{bmatrix}
		\hat{b}_{11} & \hat{b}_{12}\\
		\hat{b}_{21} & \hat{b}_{22}
	\end{bmatrix} = \bs{S}\begin{bmatrix}
		1 & \frac{\hat{a}_{12}}{\hat{a}_{22}}\\
		\frac{\hat{a}_{21}}{\hat{a}_{11}} & 1
	\end{bmatrix}\begin{bmatrix}
	\hat{a}_{11} & 0\\
	0 & \hat{a}_{22}
	\end{bmatrix}
	\label{eq:2.2}
\end{equation}

By taking the inverse of the diagonal matrix on the right-hand side of \eqref{eq:2.2}, we obtain the conventionally measured ratios. 
\begin{equation}
	\begin{bmatrix}
		\frac{\hat{b}_{11}}{\hat{a}_{11}} & \frac{\hat{b}_{12}}{\hat{a}_{22}}\\
		\frac{\hat{b}_{21}}{\hat{a}_{11}} & \frac{\hat{b}_{22}}{\hat{a}_{22}}
	\end{bmatrix} = \bs{S}\begin{bmatrix}
		1 & \frac{\hat{a}_{12}}{\hat{a}_{22}}\\
		\frac{\hat{a}_{21}}{\hat{a}_{11}} & 1
	\end{bmatrix}
	\label{eq:2.3}
\end{equation}

If we define the ratios on the left-hand side of \eqref{eq:2.3} as the measured S-parameters, we can then rewrite the remaining ratios on the right-hand side as follows:
\begin{equation}
	\begin{bmatrix}
		\overbar{S}_{11} & \overbar{S}_{12}\\
		\overbar{S}_{21} & \overbar{S}_{22}
	\end{bmatrix} = \bs{S}\begin{bmatrix}
		1 & \overbar{S}_{12}\Gamma_{12}\\
		\overbar{S}_{21}\Gamma_{21} & 1
	\end{bmatrix}
	\label{eq:2.4}
\end{equation}
where $\overline{S}_{ij}$ represents the measured S-parameters and $\Gamma_{ij}$ represents the switch terms of the VNA:
\begin{equation}
	\overbar{S}_{ij} = \frac{\hat{b}_{ij}}{\hat{a}_{jj}}, \qquad \Gamma_{ij} = \frac{\hat{a}_{ij}}{\hat{b}_{ij}}
	\label{eq:2.5}
\end{equation}

The switch terms are formed by the ratios of the receivers of the non-driving port. Therefore, they are independent of the measured DUT, as any influence introduced by the DUT will be seen equally by both waves $\hat{a}_{ij}$ and $\hat{b}_{ij}$. In general, the switch term corrected S-parameters are given as follows:
\begin{equation}
	\bs{S} = \begin{bmatrix}
		\overbar{S}_{11} & \overbar{S}_{12}\\
		\overbar{S}_{21} & \overbar{S}_{22}
	\end{bmatrix}\begin{bmatrix}
	1 & \overbar{S}_{12}\Gamma_{12}\\
	\overbar{S}_{21}\Gamma_{21} & 1
	\end{bmatrix}^{-1}
	\label{eq:2.6}
\end{equation}

In the special case where the measured two-port device is transmissionless, the switch terms $\Gamma_{ij}$ do not influence the measurements as $\overline{S}_{21}=\overline{S}_{12}=0$.

Using a four-sampler VNA, we can directly measure $\Gamma_{ij}$ by connecting any transmissive device and calculating the ratios according to the definition in \eqref{eq:2.5}. In contrast, a three-sampler VNA can only measure $\overline{S}_{ij}$. Therefore, it is advantageous for three-sampler VNAs to find a way to measure $\Gamma_{ij}$ without measuring the waves $\hat{a}_{12}$ and $\hat{a}_{21}$, and using only $\overline{S}_{ij}$ measurements.

\subsection{Proposed indirect measurement of the switch terms}

\begin{figure}[th!]
	\centering
	\includegraphics[width=1\linewidth]{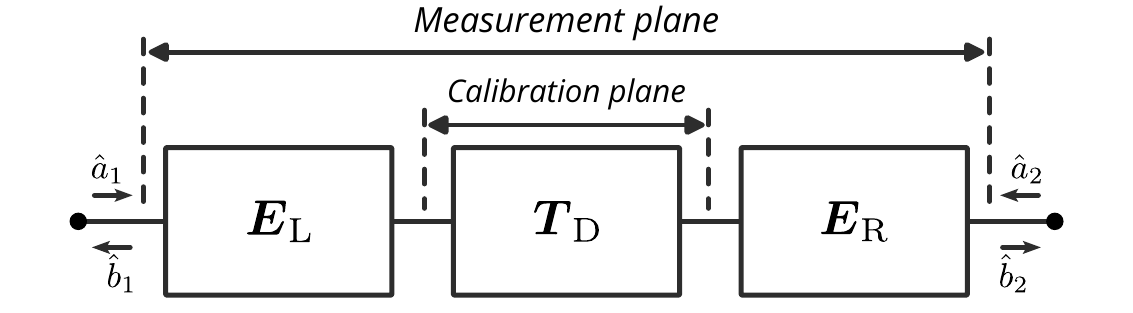}
	\caption{Two-port VNA error box model.}
	\label{fig:2.1}
\end{figure}

Fig.~\ref{fig:2.1} shows the error box model of a two-port VNA. Using T-parameters, the measured DUT is given in terms of wave-parameter as follows \cite{Rumiantsev2008}:
\begin{equation}
	\begin{bmatrix}
		\hat{a}_{11} & \hat{a}_{12}\\
		\hat{b}_{11} & \hat{b}_{12}
	\end{bmatrix} = \bs{E}_\mathrm{L}\bs{T}_\mathrm{D}\bs{E}_\mathrm{R}\begin{bmatrix}
		\hat{a}_{21} & \hat{a}_{22}\\
		\hat{b}_{21} & \hat{b}_{22}
	\end{bmatrix}
	\label{eq:2.7}
\end{equation}
where $\bs{E}_\mathrm{L}$ and $\bs{E}_\mathrm{R}$ are the left and right error boxes, and $\bs{T}_\mathrm{D}$ is the actual DUT.

We split the wave-parameter matrices in \eqref{eq:2.7} into two matrices as follows:
\begin{equation}
	\begin{bmatrix}
		1 & \frac{\hat{a}_{12}}{\hat{b}_{12}}\\
		\frac{\hat{b}_{11}}{\hat{a}_{11}} & 1
	\end{bmatrix}\begin{bmatrix}
	\hat{a}_{11} & 0\\
	0 & \hat{b}_{12}
	\end{bmatrix} = \bs{E}_\mathrm{L}\bs{T}_\mathrm{D}\bs{E}_\mathrm{R}\begin{bmatrix}
		\frac{\hat{a}_{21}}{\hat{b}_{21}} & 1\\
		1 & \frac{\hat{b}_{22}}{\hat{a}_{22}}
	\end{bmatrix}\begin{bmatrix}
	\hat{b}_{21} & 0\\
	0 & \hat{a}_{22}
	\end{bmatrix}
	\label{eq:2.8}
\end{equation}

The above expression can be simplified by multiplying the inverse of the diagonal matrix at the right-hand side. This step reduces all wave parameters into ratios as follows:
\begin{equation}
	\begin{bmatrix}
		1 & \frac{\hat{a}_{12}}{\hat{b}_{12}}\\
		\frac{\hat{b}_{11}}{\hat{a}_{11}} & 1
	\end{bmatrix}\begin{bmatrix}
		\frac{\hat{a}_{11}}{\hat{b}_{21}} & 0\\
		0 & \frac{\hat{b}_{12}}{\hat{a}_{22}}
	\end{bmatrix} = \bs{E}_\mathrm{L}\bs{T}_\mathrm{D}\bs{E}_\mathrm{R}\begin{bmatrix}
		\frac{\hat{a}_{21}}{\hat{b}_{21}} & 1\\
		1 & \frac{\hat{b}_{22}}{\hat{a}_{22}}
	\end{bmatrix}
	\label{eq:2.9}
\end{equation}

The final simplification is to replace the ratios with the definitions established in \eqref{eq:2.5}. The rearranged expression is presented in \eqref{eq:2.10}.
\begin{equation}
	\begin{bmatrix}
		1 & \Gamma_{12}\\
		\overbar{S}_{11} & 1
	\end{bmatrix}\begin{bmatrix}
		1/\overbar{S}_{21} & 0\\
		0 & \overbar{S}_{12}
	\end{bmatrix} = \bs{E}_\mathrm{L}\bs{T}_\mathrm{D}\bs{E}_\mathrm{R}\begin{bmatrix}
		\Gamma_{21} & 1\\
		1 & \overbar{S}_{22}
	\end{bmatrix}
	\label{eq:2.10}
\end{equation}

Our goal is to extract $\Gamma_{21}$ and $\Gamma_{12}$ without prior knowledge of the error boxes or the DUT. We can do this by assuming that the DUT is a reciprocal device, i.e., $\mathrm{det}\left(\bs{T}_\mathrm{D}\right)=1$. By applying the determinate operator to \eqref{eq:2.10} and using the property that $\mathrm{det}\left(\bs{A}\bs{B}\right)=\mathrm{det}\left(\bs{A}\right)\mathrm{det}\left(\bs{B}\right)$, we can derive the following:
\begin{equation}
	(1-\overbar{S}_{11}\Gamma_{12})\frac{\overbar{S}_{12}}{\overbar{S}_{21}} = \underbrace{\mathrm{det}\left(\bs{E}_\mathrm{L}\right)\mathrm{det}\left(\bs{E}_\mathrm{R}\right)}_{=c \text{ (constant)}}(\Gamma_{21}\overbar{S}_{22}-1)
	\label{eq:2.11}
\end{equation}

The above expression can be simplified as follows:
\begin{equation}
	\frac{\overbar{S}_{12}}{\overbar{S}_{21}}-\overbar{S}_{11}\frac{\overbar{S}_{12}}{\overbar{S}_{21}}\Gamma_{12} -\overbar{S}_{22}c\Gamma_{21} + c = 0 
	\label{eq:2.12}
\end{equation}

From \eqref{eq:2.12}, we can recognize that we have a linear equation in three unknowns: $\Gamma_{12}$, $c\Gamma_{21}$, and $c$. Therefore, if we measure at least three unique transmissive reciprocal devices, we can solve for these unknowns by solving the following linear system of equations:
\begin{equation}
	\underbrace{\begin{bmatrix}
		-\overbar{S}_{11}^{(1)}\frac{\overbar{S}_{12}^{(1)}}{\overbar{S}_{21}^{(1)}} & -\overbar{S}_{22}^{(1)} & 1 & \frac{\overbar{S}_{12}^{(1)}}{\overbar{S}_{21}^{(1)}}\\
		\vdots & \vdots & \vdots & \vdots\\
		-\overbar{S}_{11}^{(M)}\frac{\overbar{S}_{12}^{(M)}}{\overbar{S}_{21}^{(M)}} & -\overbar{S}_{22}^{(M)} & 1 & \frac{\overbar{S}_{12}^{(M)}}{\overbar{S}_{21}^{(M)}}
	\end{bmatrix}}_{\bs{H}}\begin{bmatrix}
	\Gamma_{12}\\
	c\Gamma_{21}\\
	c\\
	1
	\end{bmatrix} = \bs{0}
	\label{eq:2.13}
\end{equation}
where $M\geq 3$ denotes the number of measured reciprocal devices. At least three distinct measurements are required to determine the unknowns, since the system matrix must have a rank of 3 to be solvable.

The solution for the linear system of equations presented in \eqref{eq:2.13} can be obtained by finding the nullspace of the matrix $\bs{H}$. However, the accuracy of the solution is dependent on the uniqueness of the reciprocal standards. This uniqueness can be quantified using the condition number \cite{Higham1995}, defined as follows:
\begin{equation}
	\kappa(\bs{H}) = \left\|\bs{H}\right\|_F\left\|\bs{H}^{+}\right\|_F = \frac{\sigma_1}{\sigma_r}
	\label{eq:2.14}
\end{equation}
where $(\cdot)^{+}$ is the pseudo inverse and $\left\|\cdot\right\|_F$ denotes the Frobenius norm. The values $\sigma_1$ and $\sigma_r$ correspond to the largest and smallest non-zero singular values (in decreasing order), respectively, which are obtained from the singular value decomposition (SVD). In order for $\bs{H}$ to be solvable,  it requires it to have a rank of 3, hence $\sigma_r = \sigma_3$. The condition number is a relative metric for measuring the sensitivity of the solutions of a linear system of equations. The equations become more sensitive as the condition number increases. The minimum value of the condition number is one. This occurs when all the singular values are equal.

To find a solution for \eqref{eq:2.13}, it is necessary to compute the nullspace of $\bs{H}$, which can be determined by applying the SVD \cite{Strang1993}. The nullspace solution is represented by the right singular vector that corresponds to the zero singular value. The SVD of $\bs{H}$ is given as follows:
\begin{equation}
	\bs{H} = \bs{U}\bs{\Sigma}\bs{V}^H = \sum_{i=1}^{4} \sigma_i\bs{u}_i\bs{v}_i^{H}
	\label{eq:2.15}
\end{equation}
where $(\cdot)^H$ is the Hermitian transpose, $\bs{u}_i$ and $\bs{v}_i$ are the $i$th left and right singular vectors, respectively, of the corresponding singular value $\sigma_i$. The matrices $\bs{U}$ and $\bs{V}$ hold the left and right singular vectors, respectively, while the diagonal matrix $\bs{\Sigma}$ holds the singular values. It is worth noting that since $\bs{H}$ has four columns, the SVD is bounded by at most four terms.
 
Therefore, the solution for the nullspace corresponds to the fourth right singular vector, $\bs{v}_4$. As singular vectors are only unique up to a scalar multiple, a scalar ambiguity appears, as demonstrated in \eqref{eq:2.16}.
\begin{equation}
	\bs{v}_4 = \begin{bmatrix}
		v_{41} \\ v_{42} \\ v_{43} \\ v_{44}
	\end{bmatrix} = \alpha\begin{bmatrix}
		\Gamma_{12} \\ c\Gamma_{21} \\ c \\ 1
	\end{bmatrix}, \quad \forall\,\alpha \neq 0
	\label{eq:2.16}
\end{equation}

To uniquely solve for the switch terms, we can take the ratio of the elements of the nullspace vector and eliminate the scalar ambiguity. This can be written as follows:
\begin{equation}
	\Gamma_{12} = \frac{v_{41}}{v_{44}}, \qquad \Gamma_{21} = \frac{v_{42}}{v_{43}}.
	\label{eq:2.17}
\end{equation}

\section{Experiment}
\label{sec:3}

The experiment consisted of two parts. In the first step, we tested our proposed method for extracting switch terms using only three of the four available receivers on a VNA. We compared the obtained results to the switch terms computed directly using the fourth receiver. In the second step, we performed a multiline TRL calibration using both switch terms computed directly and indirectly. The results were compared by calibrating a stepped impedance line.

The R\&S ZVA, a four-sampler VNA, was used in this experiment. The reciprocal devices consisted of a line standard from the multiline TRL kit ($50\,\mathrm{mm}$ line) and a series-shunt (L-circuit) of $100\,\Omega$ resistors. We measured the L-circuit twice, by flipping the ports since it is an asymmetric device ($S_{11} \neq S_{22}$). The choice for the L-circuit is twofold. First, because of its asymmetric frequency response, it offers unique rows in the system matrix in \eqref{eq:2.13}. Second, it can cover lower frequency due to the usage of resistors.

\begin{figure}[th!]
	\centering
	\subfloat[]{\includegraphics[width=.48\linewidth]{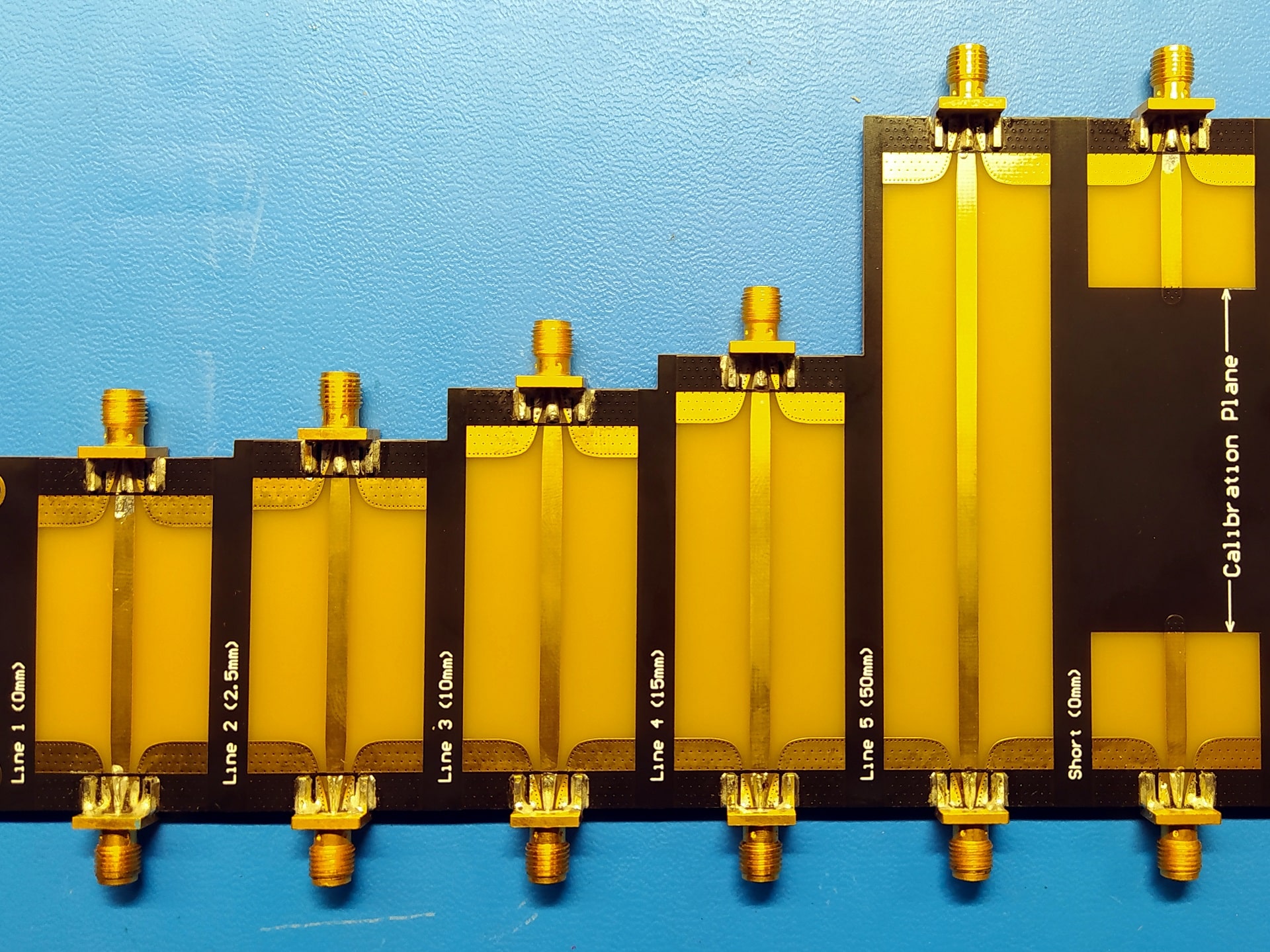}
	\label{fig:3.1a}}
	\subfloat[]{\includegraphics[width=.24\linewidth]{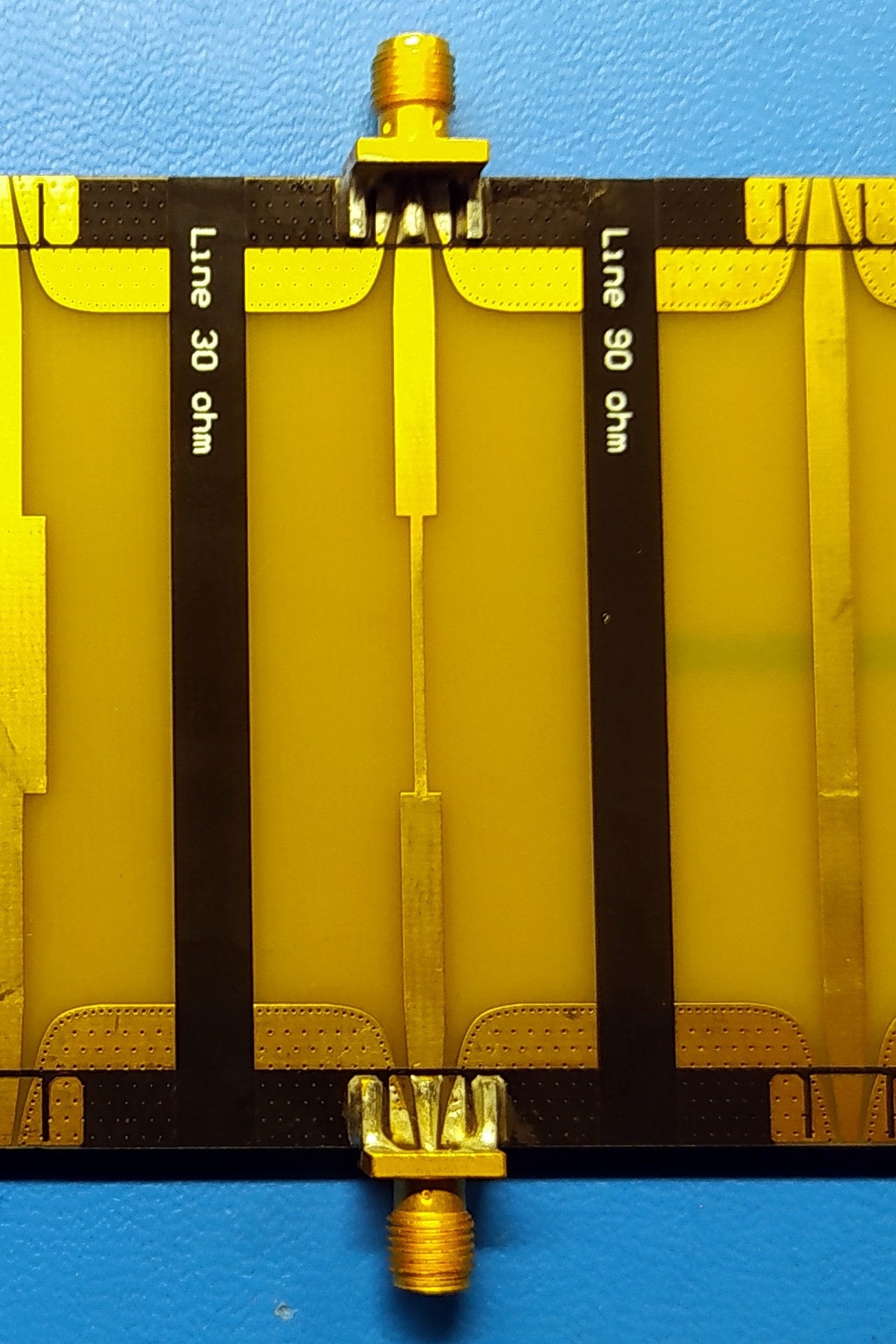}
	\label{fig:3.1b}}
	\subfloat[]{\includegraphics[width=.24\linewidth]{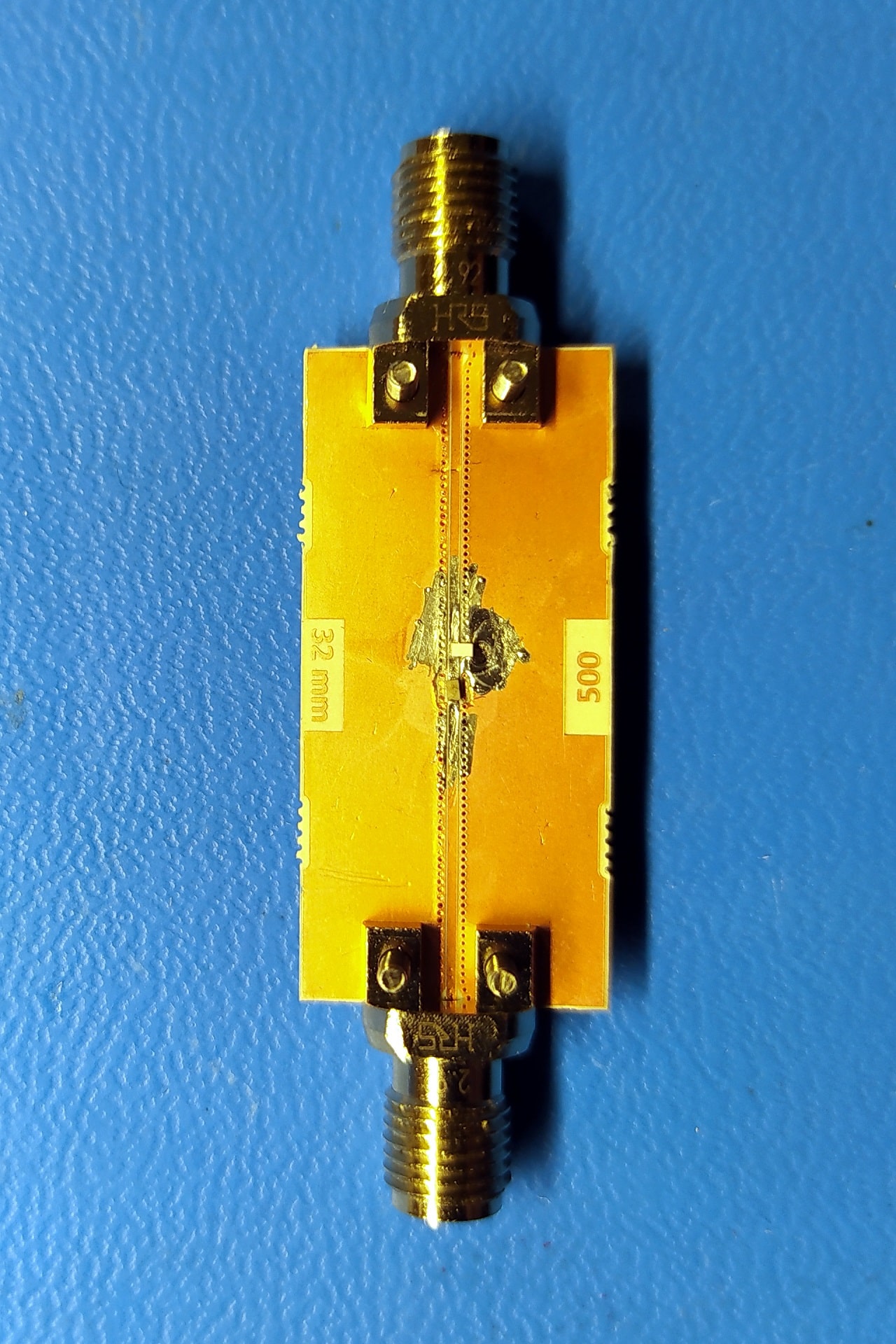}
	\label{fig:3.1c}}
	\caption{Measured structures. (a) microstrip line multiline TRL kit ($50\,\Omega$), (b) stepped impedance line ($90\,\Omega$), and (c) series-shunt $100\,\Omega$ circuit.}
	\label{fig:3.1}
\end{figure}

For the multiline TRL kit, we implemented microstrip lines on an FR4 substrate with a trace width of 3\,mm and a substrate height of 1.55\,mm. The lengths of the lines (referenced to the first line) are as follows: $\{0,2.5,10,15,50\}\,\mathrm{mm}$, with the reflect standard implemented as a short. The standards are shown in Fig.~\ref{fig:3.1}.

\begin{figure}[th!]
	\centering
	\includegraphics[width=1\linewidth]{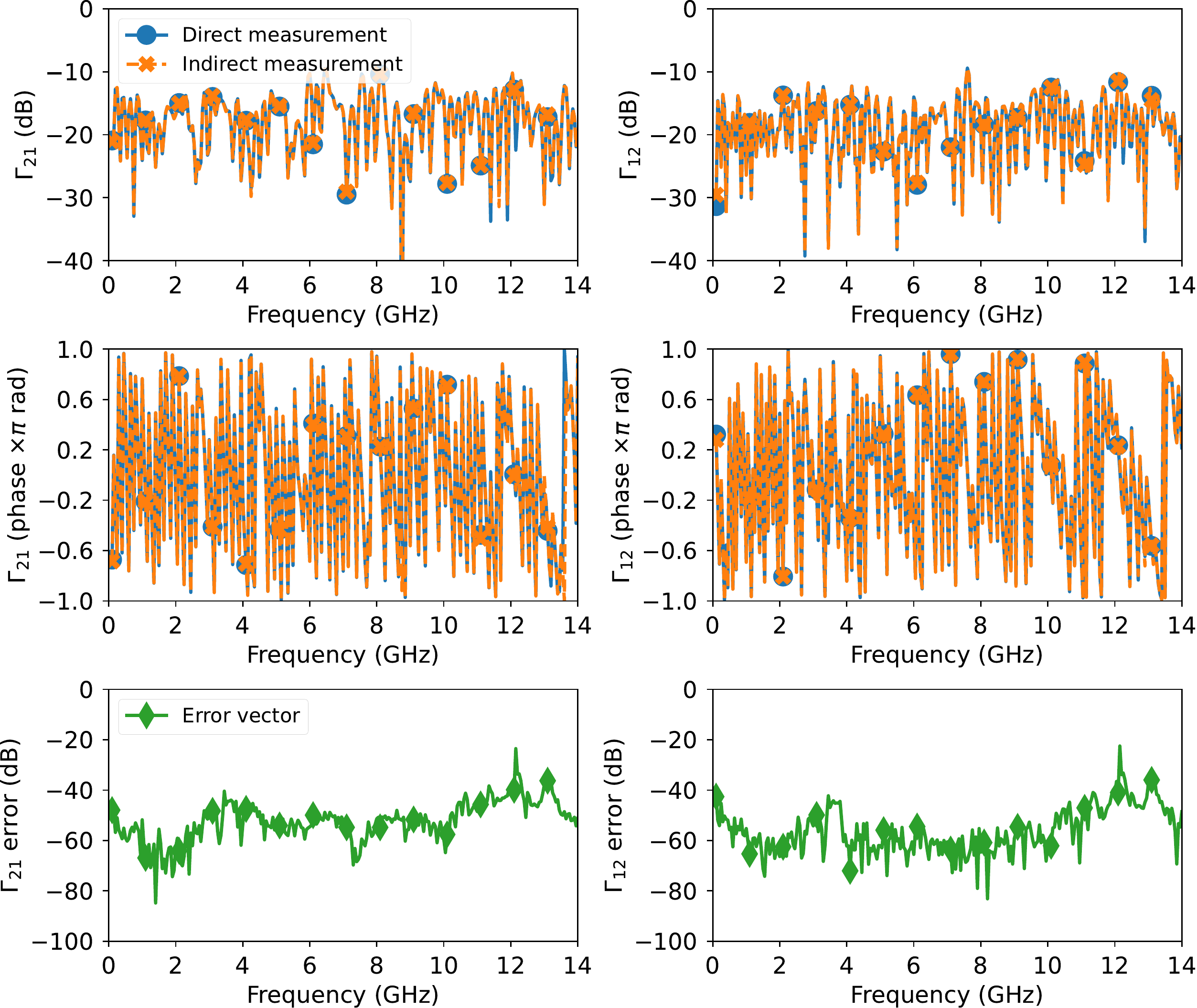}
	\caption{Comparison of direct and indirect measurements of the switch terms.}
	\label{fig:3.2}
\end{figure}

\begin{figure}[th!]
	\centering
	\includegraphics[width=0.9\linewidth]{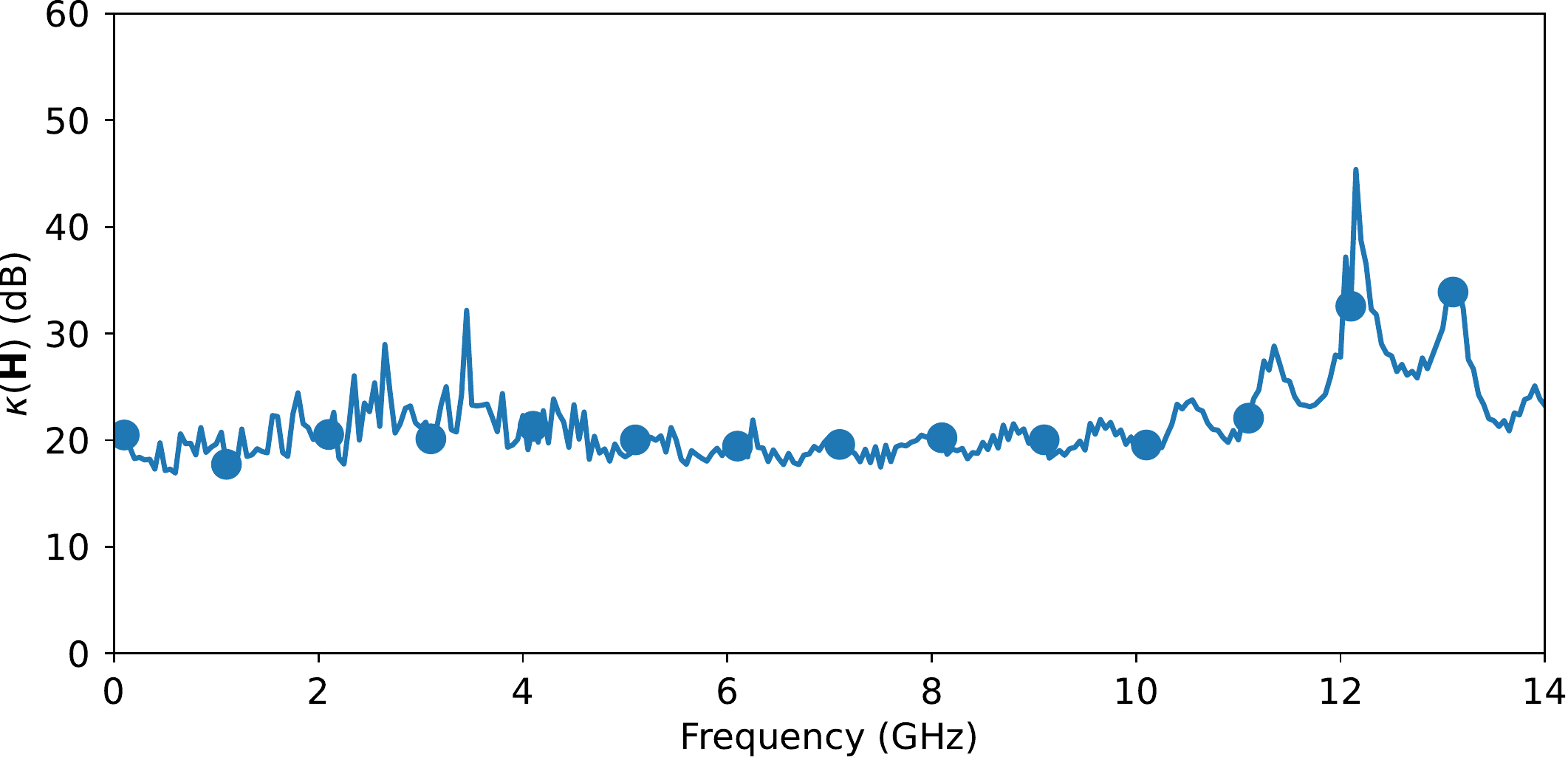}
	\caption{Condition number of the system matrix in \eqref{eq:2.13}.}
	\label{fig:3.3}
\end{figure}

To extract the switch terms, we measured the aforementioned reciprocal devices and processed the data offline using the \textit{scikit-rf} package in Python \cite{Arsenovic2022}. The results, along with those obtained by direct wave ratio computation with the fourth receiver, and the error between the two methods as defined in \eqref{eq:3.1}, are shown in Fig.~\ref{fig:3.2}. The presented results highlight that both methods overlap and demonstrate low error. However, around $12\,\mathrm{GHz}$, there is a spike in the error, which can be explained by analyzing the condition number of the system matrix, as depicted in Fig.~\ref{fig:3.3}. The condition number also demonstrates an increase in sensitivity around $12\,\mathrm{GHz}$, indicating that at these frequencies, the used standards started to exhibit a similar frequency response.
\begin{equation}
	\text{Error}_{ij}\ (\text{dB}) = 20\log_{10}\left| \Gamma_{ij}^{(\text{direct})} - \Gamma_{ij}^{(\text{indirect})} \right|
	\label{eq:3.1}
\end{equation}

\begin{figure}[th!]
	\centering
	\includegraphics[width=1\linewidth]{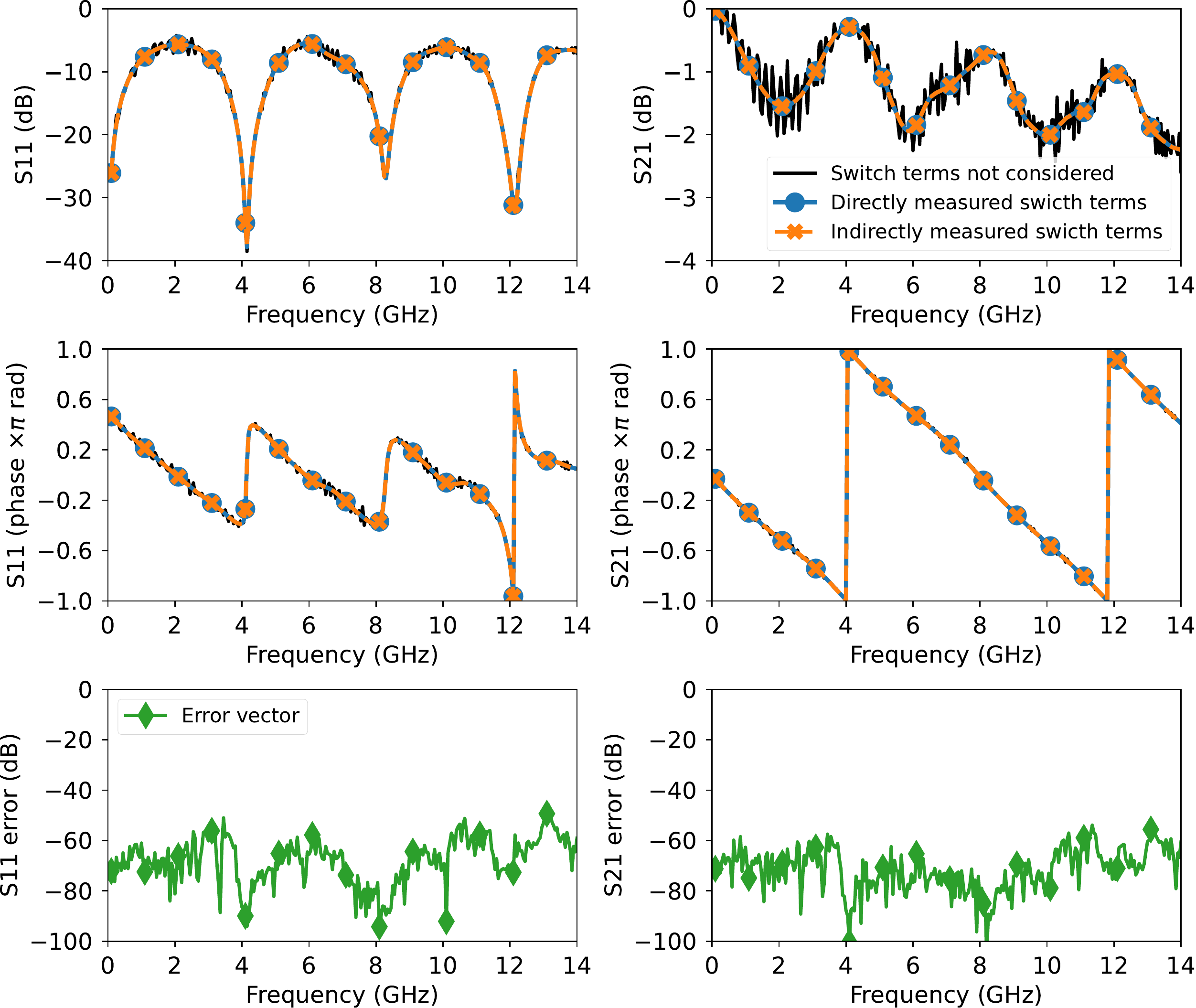}
	\caption{Results of the multiline TRL calibrated stepped impedance line.}
	\label{fig:3.4}
\end{figure}

Finally, we performed a multiline TRL calibration using the algorithm in reference \cite{Hatab2022}. In Fig.~\ref{fig:3.4}, we present the calibrated results of a stepped impedance line in different scenarios: ignoring the switch terms ($\Gamma_{21}=\Gamma_{12}=0$), directly measuring the switch terms, and indirectly measuring the switch terms, as well as the error between the methods. The results in Fig.~\ref{fig:3.4} show that both directly and indirectly measured switch terms deliver the same results, which translate as well in the error graph. However, ignoring the switch terms results in noise-like behavior on the traces.
\section{Conclusion}
\label{sec:4}
 
In this paper, we introduced an innovative method that uses only three receivers of the VNA to measure the switch terms, which requires at least three transmissive reciprocal devices. We applied this method to a four-sampler VNA, where only three of the receivers are used and compared the results with those obtained using four receivers, showing similar outcomes. Additionally, we demonstrated the successful implementation of a first-tier multiline TRL calibration, using only three receivers of the two-port VNA.

The proposed method offers a significant advantage as it does not need prior knowledge of the reciprocal devices. For instance, one can employ electronically controlled resistors at the test ports to obtain the switch terms quickly. This approach is particularly useful for error box calibration methods in multiport VNAs, allowing $N+1$ samplers to replace a dual reflectometer architecture with $2N$ samplers, where $N$ is the number of ports. This simplification considerably reduces the complexity and cost of the device.




\bibliographystyle{IEEEtran}
\bibliography{References/references.bib}

\end{document}